# From proof of concept to practice: the next decade of artificial intelligence in metasurface photonics

Sensong An[1,*], Haoyue Jiang[1]

[1]University of North Texas

*Correspondence: sensong.an@unt.edu

## Summary

Over the past decade, artificial intelligence has learned to predict electromagnetic responses rapidly and generate high-performance metasurface designs. However, one question remains open: compared with conventional design workflows that contain no learned model, has it made the complete design workflow faster, and are the resulting devices better? We argue that the answer depends on the task. Reading 100 highly cited optical studies in depth, we distinguish the design tasks in which artificial intelligence is irreplaceable, clearly advantageous, or without clear advantage. In our view, the key to the next decade is not more successes in isolated tasks but making each success a foundation for the field, through shifts such as shared data, trained models, and physics written into learning. Whether artificial intelligence leaves lasting value in this field will depend on whether its benefits can be verified, its methods transferred, and its designs corrected by experiment.

## 1. Introduction

Artificial intelligence has changed the research paradigm of many disciplines and has succeeded in specific directions. For example, in structural biology and materials science, AlphaFold solved the decades-old problem of protein-structure prediction and predicts structures accurately and at scale [1]; GNoME increased the number of computationally stable inorganic crystals by almost an order of magnitude [2]; and A-Lab joined computational target selection, robotic synthesis and measurement-guided iteration into a single automated cycle [3]. These successes share two preconditions. First, the goal is well defined and common to the field: predicting a protein structure or finding a stable crystal is a problem the whole community recognizes, so methods can be compared on the same benchmark and results accumulate. Second, a public data foundation was already in place. AlphaFold drew on five decades of the Protein Data Bank and a long-running community assessment [4,5], and materials discovery has public databases and benchmark suites of its own [6,7]. In these fields, artificial intelligence has become the default starting point. The situation in metasurface photonics is different. Although research here has grown rapidly since 2019 [8] (Fig. 1a), spanning learned structure–response surrogates [9–11], generative inverse design [12–15], end-to-end optical–computational co-design [16,17], and agentic workflows that coordinate modeling and inverse-design tools [18], artificial intelligence remains one option among others such as brute-force search and topology optimization, rather than a step that every design goes through. Both preconditions are less firmly in place here: 1) Design targets range over phase profiles, spectra, polarization, point-spread functions, and downstream task accuracy, and design spaces and evaluation criteria differ from study to study, so results do not accumulate; 2) Models and methods have accumulated quickly, but the field has not yet standardized shared datasets or data formats.

Even so, in this field, AI has delivered results in specific tasks that are hard to reach with non-AI workflows: a full-color three-dimensional waveguide holographic display built on a propagation model calibrated against measurements [19], joint optimization of the optics and the computation against final image quality [16], and inverse design that accounts for coupling between neighboring elements and extends to centimeter-scale apertures [20]. In each case, the objective lies beyond the single element, where conventional workflows lack a model that is both accurate and fast enough. However, outside such tasks, the AI approach has not spread further, for a more immediate reason: existing methods already do the job for most design tasks. Mature solvers for Maxwell's equations exist and can be checked for accuracy, so unit cells and small structures have reliable reference results. Adjoint and topology optimization remain effective routes from target response to structure [21–23], and recent commercial and foundry-linked design flows follow the same route, combining pre-characterized meta-atom libraries with electromagnetic and optical-system simulation and conventional optimization [24,25]. The design objective, moreover, usually stops at the phase or spectral response of the subwavelength element itself, and at that level parameter sweeps and conventional optimization can build a library sufficient for selection. AI therefore competes with methods that already work well, and it more often improves efficiency than capability.

To check these observations against the field as a whole, we did two things. First, we searched OpenAlex for work published between 2019 and 2026 whose title or abstract mentions both metasurfaces, metamaterials, or metalenses and machine learning, deep learning, or neural networks, restricted to the optical regime: about 1,400 papers in all. We labeled how each of these papers with an abstract was validated, which gives an estimate for the field as a whole: about 64% report simulation only, and the remaining 36% include experimental results. Second, we took about 15 highly cited optical papers published each year (100 in all) for a detailed tally, extracting information from the full text of each and labeling it against a common standard: AI-enabled workflows were separated from non-AI ones (Box 1), and we recorded how the data were acquired, what the model does, how the work was validated, what the computation cost was, and what baseline it was compared against. In this set, 27% compare against a conventional workflow and 18% evaluate both workflows quantitatively; 34% record how long data generation took and 17% reuse existing data; half include measurements (above the field-wide 36%) and 15% feed those measurements back into design (Fig. 1b and 1c). Whether AI makes metasurface design faster, or the resulting devices better, therefore cannot be read directly from most of the literature. A firmer conclusion would require comparing complete workflows, with data generation, training, design, and validation included, and separating one-time costs from those repaid over repeated use.

Based on the analysis above, together with our own work on metasurface inverse design [14,26], coupling-aware modeling [27], and metalens image restoration [28], we see progress in this field as still scattered across individual projects. A model trained for one set of design objectives and physical assumptions, including how coupling is treated, usually has to be re-simulated, retrained, and revalidated before it can serve a new task; data formats built around a specific task also limit reuse, so simulations that consume a great deal of computing time rarely serve later work. Fabrication and measurement, moreover, usually stop at validation, and the deviations they reveal are not used to update models or to guide the next round of design. If data, models, and experimental records were shared in a form that preserves information useful across tasks and carries feedback from fabricated devices, later studies could build on what has already been done. This Perspective argues for reusable, validated design systems that put AI to work on practical tasks. We propose five directions: 1) combining explicit physics with learning; 2) building shared data resources and reusable models; 3) feeding fabrication and measurement back into design; 4) optimizing for final task performance; and 5) accounting for coupling between elements where local periodic models no longer hold (Fig. 2). Section 2 reviews the capabilities built so far. Section 3 sets out a three-tier framework for judging the tasks in which artificial intelligence is irreplaceable, clearly advantageous, or without clear advantage. Section 4 discusses the five shifts, the evidence behind them, and the conditions for putting them into practice.

### Box 1. Terminology and the scope of our statistics

#### What counts as AI

**Learned model.** A trained model whose parameters carry no physical meaning of their own.

**AI-enabled workflow.** A workflow in which a learned model contributes to modelling, design, control, calibration or readout.

**Non-AI workflow.** A workflow with no learned model. Adjoint, topology, evolutionary and sweep methods, lookup tables and fitted physical models are non-AI.

**Differentiable solver.** A solver written so that gradients pass through it. Not AI by itself.

#### How designs are posed and checked

**Physics-informed.** A learned model whose architecture, loss or training incorporates physical laws.

**Local periodicity.** Computing a cell's response under periodic boundaries and reusing it where neighbors differ. A coupling-aware model conditions on the neighbors instead.

**Full-wave simulation.** A numerical solution of Maxwell's equations.

**Fabrication-in-the-loop.** Measured devices guiding subsequent designs.

#### How the literature was labelled

**Experimental validation.** The study fabricated and measured a device.

**Reuse of existing data.** A dataset released by other work enters the study's workflow.

**Complete-workflow comparison.** A full AI route, data generation and training included, against a full non-AI route for the same task. Comparisons of a single inference with a single solve do not count.

**Cost reporting.** Hardware, data-acquisition time and training time.

#### Where the numbers come from

**Optical metasurface literature.** Optical works published in 2019-2026 whose title or abstract mentions metasurfaces, metamaterials or metalenses, excluding microwave, terahertz, acoustic and similar work: 25,349 in all (OpenAlex).

**AI-related literature.** The roughly 1,400 of these whose title or abstract also mentions deep learning, machine learning or neural networks. They give the publication counts and the AI share (Fig. 1a).

**Abstract-labelled works.** The 1,138 of these with an abstract, labelled by a language model (Claude) under fixed rules for whether they include experiments. They give the field-wide experimental share (about 36%).

**Full-text studies.** About 15 highly cited studies per year, 100 in all, labelled from the full text and checked by the authors. They give the remaining shares and counts (Fig. 1b and 1c, Fig. 2).

*Every number with its source: Supplementary Table S2. Definitions and methods: Supplementary Table S1 and Supplementary Methods.*

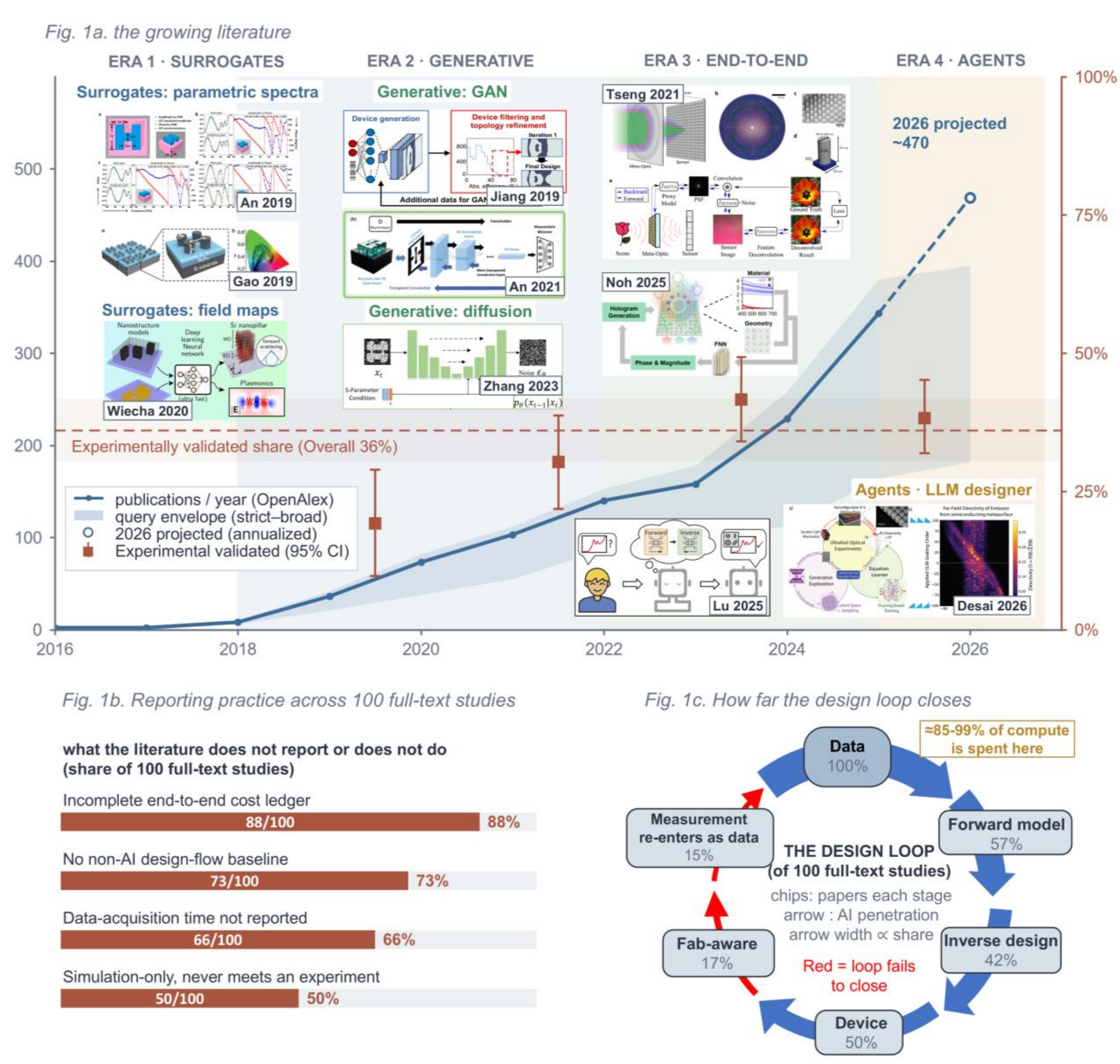

**Figure 1. Research growth, evidence gaps and experimental feedback in AI-enabled optical metasurface design. a,** Annual OpenAlex counts of AI-enabled optical metasurface publications (blue, left axis; optical restriction as in Supplementary Methods 5); shading shows variation across queries, and the dashed segment and open marker indicate the annualized 2026 estimate. Experimental-validation shares (right axis) are calibrated field-wide estimates (all 1,138 optical works with a usable abstract label, not a sample; squares; two-year bins; 95% confidence intervals). The horizontal dashed line marks the overall field-wide estimate. Insets show representative capabilities and are adapted from [26], [29], [11], [13], [14], [15], [16], [30], [18], and [31]; era bands indicate shifts in emphasis among coexisting approaches. **b,** Gaps in cost reporting, non-AI workflow comparison and experimental validation. Cost completeness covers hardware, data-acquisition time and training time. Workflow comparisons exclude inference-only comparisons. **c,** Schematic cycle linking computational design, fabrication, measurement and subsequent

updates. Percentages indicate the prevalence of each activity; categories overlap, and arrows denote information flow rather than transition rates. Fabrication-aware design is distinguished from measurement-informed updates; dashed links highlight opportunities for stronger integration. The full-text set contains 100 optical studies (Box 1). Definitions, statistical methods and indicator-specific denominators are given in the Supplementary Methods, and every number in the figure is listed with its source in Supplementary Table S2.

## 2. The first decade: what got built

*These four lines of work overlap rather than succeed one another.*

**Learned surrogates (since 2017).** Researchers first used AI to build learned surrogates that could rapidly predict the electromagnetic response of metasurface structures. A neural network was trained on simulated structure–response data and, once trained, could predict the response of a new structure far faster than a full-wave simulation. This speed requires an up-front investment: the training data must first be generated by full-wave simulation. In our corpus, 54% of studies build their training sets through predetermined sweeps or random sampling (Supplementary Table S1). Neither approach adapts where it samples based on earlier results, so the number of simulations required is often large, and generating the training set typically takes longer than training the network and accounts for most of the total effort. In return, a trained model can evaluate large numbers of candidate structures in a short time, making it possible to search a design space for structures that meet a target response, and so surrogates could also be used for inverse design. Representative early work includes tandem networks, which pair a surrogate with an inverse network to handle the one-to-many nature of inverse design [32]; structure–response models and inverse-design methods for plasmonic nanostructures, multilayer nanoparticles and all-dielectric meta-atoms [26,33,34]; and bidirectional models for structural color and chiral metamaterials [29,35]. Surrogates have since become a standard tool, used in more than half (57%) of the studies in our corpus. The central questions for this line of work remain the same: whether surrogates stay reliable for structures that differ substantially from their training data, whether the cost of generating data can be recovered through reuse, and how to move beyond single-period unit cells.

**Generative inverse design (since 2018).** Tandem networks built on trained surrogates can handle inverse design for simple meta-atoms with few parameters, but for structures with complex shapes and many degrees of design freedom they struggle to generate feasible geometries directly. Researchers therefore turned to generative models, which learn a probabilistic mapping between target responses and geometries and can thus return multiple distinct candidate structures for the same target response. The first such methods were based on generative adversarial networks (GANs), which generated candidate patterns for prescribed spectra [12], diverse metagratings for specified wavelengths and deflection angles [13], and geometrically distinct structures with similar electromagnetic responses [14]. Because GANs rely on adversarial training between a generator and a discriminator, training is often unstable and the outputs tend to collapse toward a few similar designs. MetaDiffusion later brought diffusion models to the inverse design of free-form meta-atoms, generating structures by denoising step by step from random noise, which made training more stable and the candidates more diverse [15]. Because a generative model returns a set of candidates, which of them actually meet the target still has to be checked; existing studies have done so with FEM simulation [12], a learned forward model [15], or RCWA evaluation followed by topology optimization [13].

**End-to-end differentiable design (since 2021).** Surrogate and generative models were mostly designed for the response of the metasurface unit itself. In imaging, holography and similar applications, however, a design is judged by the final output of the whole system, such as the quality of a reconstructed image or the accuracy of a downstream task. Researchers therefore wrote the entire chain from the metasurface to the final output as a differentiable model. The optical frontend describes how light passes through the metasurface to the detector; the computational backend turns the detected signal into an image or a task result. Gradients can then propagate from the final objective back to the optical structure, so that the optical design directly serves that objective. Since 2021, this approach has been used in imaging, holography, sensing and optical computing. These studies fall broadly into two groups, depending on where the learned model sits in the chain. In the first, the learned model forms the computational backend. A neural network turns the signal recorded by the detector into the final image, for example by restoring the blurred image captured through a metalens, and is optimized together with the metasurface as part of the system from the design stage onward [16,17]. In the second, the learned model sits in the optical frontend and replaces simulation: the network learns from data to predict quickly how light responds to the metasurface, and during design, gradients pass through it back to the design variables [30,36,37].

Every optimization loop above contains a learned component. End-to-end differentiability alone, however, does not make a workflow AI-enabled. In compressed-sensing imaging [38], thermal imaging [39] and multidimensional holography [40], gradients pass through explicit reconstruction algorithms or physical propagation models, and no model fitted to data appears in the loop. Under the definition in Box 1, these are end-to-end but non-AI workflows.

**Agents and reusable models (since 2025).** In the three lines of AI-enabled work above, the workflow is assembled in advance by the researchers, with fixed rules for how the steps connect and when each tool is called, and the models are usually trained for a single task, so a new task requires new data and retraining. Since 2025, researchers have begun to look for more efficient approaches. At the workflow level, design is increasingly orchestrated by agents: an agentic framework coordinated simulation, surrogate modelling, optimization and inverse-design tools [18], and a self-driving laboratory extended automated decision-making into experiment, using each round of measurements from a reconfigurable light-emitting metasurface to choose the next test [31]. At the model level, researchers have begun to build reusable models and frameworks: MetasurfaceViT used physics-informed data augmentation to expand its Jones-matrix dataset to more than 60 million samples, was pretrained once on that dataset, and could then be fine-tuned for inverse design across wavelengths, polarizations and applications [41]; MAPS integrated data generation, training, benchmarking and inverse design into a configurable open-source framework spanning multiple classes of photonic devices [42]. These studies allow workflows to adapt to the task at hand and models to be reused across design problems without retraining from scratch each time.

**The balance sheet.** Over the past decade, the field has developed learned surrogates, generative inverse design, end-to-end differentiable design, and agents and reusable models, and has trained a generation of researchers versed in both electromagnetics and machine learning. Yet AI has not become a routine step in metasurface design: in the OpenAlex search described in the Introduction, studies involving AI account for about 6% of the optical metasurface and metamaterial literature, rising to about 8% in the past two years (Supplementary Methods). As noted in the Introduction, most studies do not compare their results quantitatively with a conventional workflow, so the literature does not yet show on which tasks AI outperforms conventional methods. Existing data are also rarely reused. Experimental validation is also limited: the share of studies with experimental results rose from about 20% in 2019–2020 to about 40% in recent years but remains below half (Fig. 1a). Most studies generate their own data and stop at simulation, so their results are hard to compare with one another and hard to carry over to real devices. The decade therefore leaves two questions: 1) when does AI offer a clear advantage over conventional methods, and when are conventional methods already sufficient? 2) how must research practice change for the value of AI to extend beyond isolated demonstrations? Section 3 addresses the first question; Section 4 the second.

## 3. When is AI enabling, efficient, or unnecessary?

When does AI offer a clear advantage over conventional methods, and when are conventional methods already sufficient? The answer starts with forward prediction, since every design workflow, with or without AI, depends on it. Conventional optimization and search call it repeatedly; the training data for learned surrogates are mostly generated with it, and pure inverse networks that map responses directly to structures are no exception; and design results must ultimately be validated with it. The role AI can play in a task therefore depends on two things: first, whether a non-AI forward model (a solver or an explicit physical model) exists that is both accurate and fast enough; second, if one does, whether the cost of the workflow is dominated by many repeated, expensive simulations that a learned model can reduce at lower total cost. On this basis, we divide design tasks into three categories (Table 1): when no such model exists, AI is enabling (T1); when such a model exists but repeated simulation dominates the cost, AI is efficient (T2); when such a model exists and few repeated simulations are needed, AI is unnecessary (T3). This division assumes current solvers, computational resources and patterns of use, and its boundaries shift as these conditions change, as we discuss at the end of this section.

**Enabling (T1).** AI is enabling when a task would be difficult to accomplish without it. Such tasks lack a non-AI forward model that is both accurate and fast enough, usually for one of two reasons. First, some step in the workflow cannot be written as an accurate explicit model and can only be learned from data. In one case, the actual hardware departs from the ideal physical model: fabrication deviations, assembly alignment, material parameters and sensor response can be determined only by measurement. In another, the step is itself defined by the statistics of data: the priors that image reconstruction relies on, for example, cannot be derived from first principles. Here, learned models allow this step to enter the design loop. A waveguide holography model calibrated with measured data [19] and image-reconstruction networks trained jointly with metasurface optics [16,17,43] are examples. Conversely, when reconstruction can be written directly from physical laws, learning is no longer necessary: in thermal imaging, inversion based on Planck's law generalizes better than neural-network reconstruction [39]. Second, an accurate model exists but cannot be computed at the required scale or within the required time, while the approximations that can be computed neglect effects the design depends on. Full-wave simulation is accurate, but its computational cost grows rapidly with device size; local periodicity (Box 1), diffraction theory and the angular spectrum method are fast but neglect coupling between elements. When a design depends on exactly this coupling, as in large-aperture devices and strongly coupled arrays, a learned surrogate can compress the results of full-wave simulation into a fast model. In one infrared diffuser design, full-wave simulation could be completed in a practical time only for arrays up to about 12 × 12 elements [44]. A coupling-aware model, by contrast, scaled inverse design to centimeter-scale apertures; its accuracy was checked against full-wave simulation on 24 × 24-element arrays and by measurement on devices up to 224 μm across [20]. For coupled arrays, learned surrogates also predict responses more accurately than these approximations [27,37]. What matters in evaluating these tasks is what was achieved, not how much faster it was.

**Efficient (T2).** AI is efficient when a task can be accomplished without it but AI clearly lowers the total cost. Such tasks already have a non-AI forward model that is both accurate and fast enough, and most of the cost goes into repeated simulations; the role of a learned model is to reduce them, saving more computation than data generation and training consume. The savings come from two sources. The first is amortization across tasks: the same class of forward problem must be solved for many targets. A conventional workflow must re-optimize for each new target, so its cost grows linearly with the number of targets; a learned model requires data generation and training only once, after which each new target costs little. Once the number of targets exceeds a break-even point, the AI workflow costs less in total than the conventional one. Pretrained models reused across wavelengths, polarizations and applications [41], and structural-color design that must serve many target colors [45], both rely on this amortization. The second is search within a single task, when the design space is high-dimensional and non-convex. Gradient-based local methods tend to stall in local optima, while global search requires many expensive evaluations; learned policies or generative models can concentrate evaluations in more promising regions, finding better solutions within the same budget or reaching the same result with fewer evaluations [46,47]. This advantage is conditional, however, and varies with the problem and the design space: the authors of [46] note that rankings may change once each method is fully tuned. What matters in evaluating these tasks is total cost, including data generation and training: the claim requires a complete-workflow comparison (Box 1) under a matched budget, reporting the number of uses needed to break even. As the statistics in the Introduction show, such comparisons remain rare.

**Unnecessary (T3).** AI is unnecessary when conventional methods already suffice and introducing AI would not clearly reduce the total cost. Such tasks usually meet three conditions at once, so the up-front cost of learning is hard to recover. First, the forward model is itself accurate and cheap, as in element-level problems where local periodicity holds. Second, optimization is already efficient: with few parameters, a direct sweep suffices; with many, the adjoint method obtains gradients with respect to all parameters from one forward and one adjoint simulation [21–23], so as long as a local optimum meets the design requirements, high dimensionality does not by itself favor AI. Third, the task is one-off or small-batch, with no repeated use over which to amortize up-front costs. In such tasks, the larger design freedom or search range that AI offers often goes unused, and what it adds is the cost of data generation and training. In one single-layer antireflection-coating design, for example, a free-form generative network searched a far larger space than a parameter sweep and reached only slightly lower reflection than the best swept design, but its training took a few hours, several times longer than a finer parameter sweep; a coarse sweep took only 6–7 minutes on the same 16-core processor [48]. Forward prediction is no different: for known partial differential equations, physics-informed neural networks have not yet surpassed the finite-element method in either speed or accuracy [49]. Most element-level design falls into this category, consistent with the assessment in the Introduction that existing methods already suffice at this level.

**The boundaries shift.** The division into these three categories assumes current solvers, computational resources and patterns of use, so the boundaries between them are not fixed. Consider first the boundary between T1 and the other two categories: as non-AI forward models become faster and more accurate, some T1 tasks will move to T2 or T3; conversely, as design targets extend from element responses to the device and system levels, new steps will emerge that lack such a model, and T1 will grow. Consider next the boundary between T2 and T3: when the same class of task recurs and the number of uses exceeds the break-even point, a T3 task moves to T2; when data and models can be shared and reused, up-front costs are spread across many studies and the break-even point falls. The five shifts discussed in Section 4 aim to move these boundaries. Combining physics with learning reduces the training data required, and building shared data and reusable models spreads up-front costs; both lower the threshold for T2. Feeding fabrication and measurement back into design, targeting final task performance, and accounting for coupling between elements bring into the design loop more steps that cannot be written as explicit models or cannot be computed in time, widening T1.

**Table 1. Tasks in which AI is enabling, efficient or unnecessary in metasurface design**

| Tier | Typical cases | Evidence required |
|---|---|---|
| **T1: enabling** | Hardware departs from the physical model [19]; image reconstruction [16,17,43]; large-aperture or strongly coupled arrays [20,27,37,44] | Independent confirmation by full-wave simulation or measurement |
| **T2: efficient** | Amortization across many targets [41,45]; high-dimensional non-convex search [46,47] | Complete-workflow comparison under a matched budget; number of uses needed to break even |
| **T3: unnecessary** | Element-level design [48]; problems solvable by sweeps or adjoint methods [21–23]; one-off tasks | A sweep or adjoint design that meets the specification within budget |

*The classification assumes current solvers, computational resources and patterns of use, and it shifts as solvers improve, as uses accumulate and as resources are shared (Section 4).*

## 4. Five shifts for the next decade

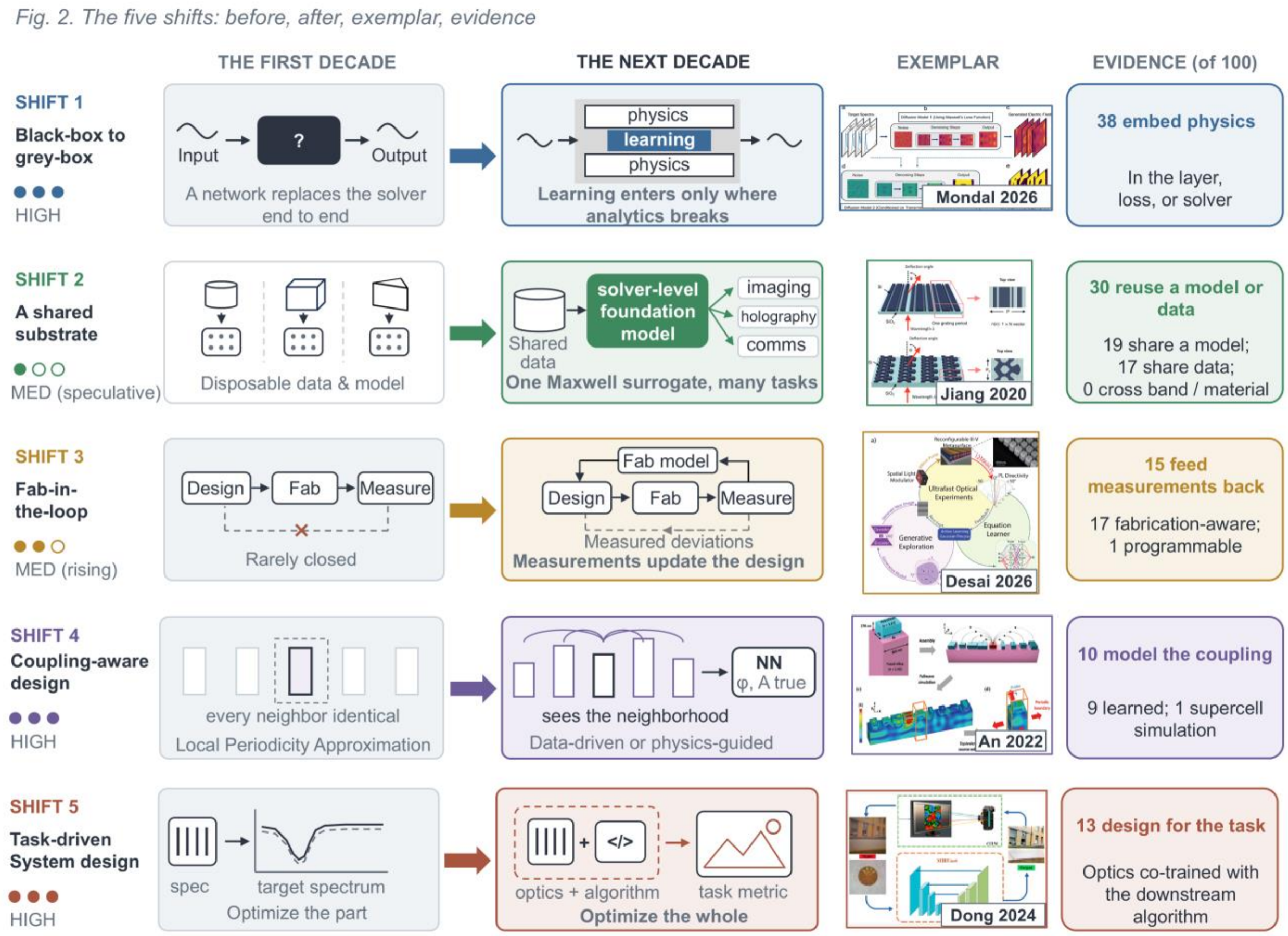

**Figure 2. Five shifts.** One row per shift of Section 4: from black-box to grey-box models; from single-use models to shared foundations; from simulation-only to fabrication-in-the-loop; from local periodicity to coupling-aware design; and from unit-cell metrics to task-driven system design. Each row shows the prior practice, the practice replacing it, a published exemplar, and the corresponding counts among the 100 full-text studies (Supplementary Table S2). Exemplars adapted from [50], [51], [31], [27], and [28].

### Shift 1: From black-box to grey-box models

In 60 of the 100 studies in our detailed tally, the model itself contains no physics: solvers only generate training data and validate results, and the network relearns from data the relationships that Maxwell's equations already provide. Such black-box models need large datasets, and their predictions become hard to trust outside the training distribution. Yet the answer is not to let a network replace the solver altogether: even PINNs, which add equation residuals to the loss, have not surpassed the finite element method in speed or accuracy on benchmark forward problems [49,52] (Section 3). A more promising direction is the grey-box model, which assigns to physics whatever can be written as equations and leaves to learning only what cannot, or what would be too expensive to compute directly, shrinking the black box to what is necessary. Physics enters such models in four main ways. First, it can be built into the network architecture: complex-valued linear layers represent slab-mode propagation, so the network structure itself mirrors how light propagates [37]. Second, it can act as a constraint in the loss, what is usually called physics-informed learning. Unlike a PINN that replaces the solver, here the equation residual only constrains the output of the learned component: MxDiffusion adds a Maxwell-equation residual to its training loss to constrain the generated electric fields [50], and a multiscale physics-informed network combines Helmholtz-equation residuals with a fit to simulated field data to retrieve a spatially varying permittivity distribution [53]. Third, a differentiable solver can sit inside the loop: GLOnet uses forward and adjoint simulations to compute device-efficiency gradients and backpropagates them to a generative network [47]. Fourth, a physical model can carry most of the description, with learning correcting only the residual: in waveguide holography, an explicit transfer function describes the known propagation, and camera measurements are used only to fit the coupler-efficiency maps and residual aberrations that the physical model misses [19]. Because known physics need not be relearned from data, grey-box models require less training data; the known part still holds outside the training distribution, and the outputs agree more closely with physical laws. This is one way to lower the threshold at which AI becomes efficient (Section 3).

### Shift 2: From single-use models to shared foundations

As noted in Section 2, most learned surrogates are trained for a single task on data generated only for that task; a new task means new simulations and retraining. This up-front investment is substantial: in two studies that report a time breakdown, simulation accounts for 85–99.8% of the combined data-acquisition and training time [9,45], and some data campaigns last weeks to months [54,55]. Remedies include active and Bayesian sampling, which select samples adaptively and so reduce the number of simulations or experiments required [31,56], and transfer learning and physics-residual pretraining, which reduce the data needed for the target task [57]. These methods, however, lower the cost of individual projects, and each project still bears its own up-front investment. As Section 3 showed, the more targets a model serves, the more thinly its up-front cost is spread; if this amortization extended beyond a single project to other projects and research groups, the break-even point would fall further. Existing work shows that this is feasible. A meta-atom dataset released with our earlier modeling study [58] was later used directly to train a diffusion model, without generating new simulation data [15]; the physics-residual pretraining mentioned above [57] used a public meta-atom dataset and needed only 1,800 target samples to transfer to absorber design; and MetaNet provides a platform for sharing photonic data, designs and code [51]. Such reuse nonetheless remains rare: only 17% of the studies in our tally reuse existing data (Introduction).

In our view, three steps would help sharing take hold. The first is a common content and format for records. Instead of a project's design metrics, a record can store the field itself, from which transmission, phase, polarization response and far-field patterns can all be derived. A record holding the normalized complex field, together with geometry, constitutive parameters, excitation, boundary conditions and solver metadata, can therefore serve different targets across projects. Such records are also well suited to training solver-level models, which learn the mapping from structure to field directly; neural operators already learn such mappings for frequency-domain devices, three-dimensional scattering, resonant structures and time-domain evolution [59–63]. The second is making data usable across bands. When material dispersion is negligible, the dimensionless Maxwell equations are scale invariant, so with geometry expressed in electrical lengths, data from one band can be used in another. When dispersion matters, recording material parameters explicitly keeps the data usable; in optical metal antennas, for example, finite skin depth invalidates simple wavelength scaling [64]. Frequency-conditioned operators can already predict fields at untrained wavelengths within the trained band [65]. The third is community mechanisms that encourage sharing. The CIF standard in crystallography, which ties a common data format to publication, provides a precedent [66]; PDEBench [67] and The Well [68] show that standardized datasets and interfaces make methods directly comparable and data reusable. With such shared foundations, each expensive full-wave simulation can serve multiple models and design targets, the up-front cost is shared across the community, and the threshold at which AI becomes efficient falls.

### Shift 3: From simulation-only to fabrication-in-the-loop

Simulation is an indispensable step in design, but the value of a design is ultimately confirmed through fabrication and measurement. In most current work, however, fabrication and measurement serve only as a one-time validation at the end of the chain: in our tally, only 17% of studies bring manufacturability into the design loop, and just 15% feed measurements back into design (Introduction; Supplementary Table S1). Deviations caused by process variation, calibration error or physics missing from the solver are recorded but rarely used to update the model or guide the next design. Two complementary approaches bring fabrication and measurement into the design loop. The first is fabrication-aware optimization, which builds known deviations into the design before fabrication: geometric rules exclude structures that violate minimum feature-size or spacing requirements [69], learned surrogates enable robustness optimization under fabrication uncertainty [70], and a learned lithography model drives optical proximity correction, yielding a measured metalens efficiency of 64%, close to the calculated 69% [71]. The second is measurement-driven redesign, which uses measured results to correct subsequent designs and addresses deviations that could not be anticipated and that the model misses: after six rounds of fabrication with spectral measurements fed back into design, a coupler reached an insertion loss of 3.24 dB, compared with a baseline of 8.8 dB [72].

Once the loop is closed, the bottleneck moves to the hand-offs between its steps: researchers still pass information manually between simulation, training, optimization and experiment. Agents can automate these hand-offs. An agentic framework already coordinates simulation, model training and inverse design [18], and a self-driving laboratory selects each next experiment from the measurements of the previous round [31] (Section 2). We see the next step as combining the two: designs iterate rapidly in simulation, and after each round of fabrication and measurement, the measured deviations are used to correct the models that the simulation relies on. Researchers set the objectives and physical constraints; agents handle the hand-offs in between.

### Shift 4: From local periodicity to coupling-aware design

Most library-based metasurface designs rely on local periodicity: the response of a single unit cell is computed under periodic boundary conditions and then used at positions in the device where the neighbours differ. This approximation makes library-based design simple and practical, but it is also a common source of deviation between the fabricated device and its design, and it is rarely quantified at the device scale. Full-wave simulation of the whole aperture avoids it, but at a cost that becomes prohibitive for electrically large devices [44] (Section 3). Of the 100 studies in our tally, only 10 actually compute or learn inter-element coupling in the design: 1 computes it with supercell simulations, and 9 capture it with learned models; another 3 weaken coupling through larger spacing or modified unit-cell structures. The key to a coupling-aware model is to condition the response of each element on its neighbours rather than on an isolated periodic cell. Such a model can predict the perturbed amplitude and phase of a target meta-atom from its neighbours [27], or the field scattered by a local arrangement of several elements [10]. Because only interactions within a finite neighbourhood are learned, a trained model can assemble the response of an entire aperture from overlapping neighbourhoods and be applied, without retraining, to much larger devices [20]. Supercell and stitching methods remain effective at manageable scales. Learned models extend to large aperiodic apertures, and their results can be checked against full-wave simulation on representative regions. Note, however, that when a structure supports laterally propagating modes, such as guided modes or surface waves, coupling extends beyond the finite neighbourhood and the accuracy of these methods declines.

### Shift 5: From unit-cell metrics to task-driven system design

Most metasurface designs still target responses at the unit-cell level, such as phase, transmission spectra, the relative phase between polarizations, or the phase at different frequencies. These quantities are only proxies for system performance. Even when every unit cell matches its target phase, the assembled metalens may not deliver the intended focal intensity and point spread function, let alone a better reconstructed image or higher recognition accuracy. Of the 100 studies in our tally, only 13 use reconstruction quality or task accuracy as the design objective. End-to-end differentiable design, introduced in Section 2, offers another route: it links the physical variables of the metasurface to the final output of the system and optimizes directly for reconstruction quality or task accuracy [16,17]. What changes is not only the objective. When optics and computation are optimized jointly, performance can be redistributed between optical encoding and computational recovery. The optical element no longer has to meet strict intermediate specifications on its own, and residual aberrations can be left to the reconstruction algorithm [28]. Nor are system-level objectives limited to imaging. The same idea has been applied to polarimetric spectral sensing [36] and optical computing [37], where the metasurface acts as an encoder in an information-processing chain rather than as a stand-alone optical element.

Targeting system metrics presupposes an accurate simulation of the system, and that is itself difficult. From unit-cell responses to the field of the whole device, and then through propagation, detection and reconstruction, the error at each step accumulates in the final metric. The earlier shifts supply the corresponding tools. Coupling-aware models assemble unit-cell responses accurately into device-level fields (Shift 4); solver-level models return field distributions quickly (Shift 2); and system models calibrated with measured data capture deviations that simulation misses [19] (Shift 3). System-level design also brings in stages that have no explicit model, such as the data priors on which image reconstruction relies. Only learning can handle these, so pushing the design objective toward the final task also widens the range in which AI is enabling (Section 3). Comparison is a further difficulty. As noted in the Introduction, the success of AI in structural biology and materials science rested on shared objectives and public benchmarks. System-level design lacks exactly this: studies differ in optical system, reconstruction method and task, so their results are hard to compare and hard to build on. Common system models and evaluation standards for typical tasks such as imaging and spectral sensing would allow different methods to be compared on the same task, and progress could then accumulate. In our view, the larger opportunity lies in starting from the task and working backwards to how the metasurface encodes information.

## 5. Outlook

The lasting value of AI in metasurface photonics will depend on what researchers can reliably build with it. The past decade showed what AI can do within individual projects: it enables designs that conventional methods cannot easily reach, and it speeds up design in many tasks. The question for the next decade is whether these gains can accumulate across projects. Section 3 asked where AI is useful and separated tasks in which AI is enabling, efficient or unnecessary. The five shifts in Section 4 describe how its value can accumulate. They are writing known physics into models, sharing data and models between groups, feeding fabrication and measurement results back into design, accounting for inter-element coupling at the device scale, and targeting the final task rather than intermediate metrics. As noted in the Introduction, the success of AI in structural biology and materials science rested on public data and common benchmarks. If the metasurface community builds a similar foundation, each study can give the next a better starting point. We envision that, in metasurface photonics, AI will no longer be a new method that has to be justified case by case, but will gradually become a reliable tool that researchers choose naturally wherever it adds value.

## Acknowledgments

We thank Juejun Hu for his insightful guidance and constructive suggestions that helped shape the ideas presented in this Perspective.

## Declaration of interests

The authors declare no competing interests. For transparency, references 14, 26, 27, 28, 58, and 69 are the authors' prior work, and the public meta-atom dataset reused in reference 15 was released by the authors.

# Supplementary Information

## From proof of concept to practice: the next decade of artificial intelligence in metasurface photonics

Sensong An[1,*], Haoyue Jiang[1]
[1]University of North Texas
*Correspondence: sensong.an@unt.edu

### Contents

## Supplementary Methods

### 1. Scope

'Optical' means metasurface and metamaterial work from the ultraviolet to the infrared (free-space wavelength up to about 30 μm, i.e. above about 10 THz). Microwave, millimetre-wave, terahertz (0.1–10 THz), acoustic and mechanical work is excluded from both the field-wide survey and the full-text set.

### 2. Full-text set

For each year from 2019 to 2025, the AI-enabled optical works of the field-wide survey (Supplementary Methods 5) were ranked by OpenAlex citation count and taken in order until 15 eligible studies were reached (2019: only 10 eligible). Eligible means primary research, optical, an engineered nanostructure, and a learned model used by the authors. Works without an obtainable full text (93), or found in full text to be non-optical (12) or without a learned model (3), were replaced by the next eligible work.

Each study was labelled from its full text, with a verbatim quote for every label, and checked by the authors.

All full-text percentages use the 100 studies as denominator. The set is not a random sample: it holds the most-cited studies and is more often experimental than the field (50% against about 36%).

### 3. Comparison with a non-AI workflow

Each study is classed as quantified (a non-AI workflow was run and compared quantitatively), run but not quantified, cited only (a non-AI baseline cited or asserted but not run), or none. Comparisons limited to per-inference time against one solver call are classed as per-inference only and not counted. 'Compare against a conventional workflow' counts quantified and run but not quantified; 'evaluate both workflows quantitatively' counts quantified.

### 4. Coupling and system-level objective

Inter-element coupling is counted when it is computed with physical methods (for example supercell simulation) or learned by a model conditioned on neighbouring elements. Studies that only reduce coupling (larger spacing, isolating structures) are reported separately; studies that only mention it are not counted.

A system-level objective is counted when design variables are optimized directly against reconstruction quality or task accuracy. Holograms judged only by intensity, and back ends trained after the metasurface was fixed, are not counted.

**5. Field-wide survey**

Queries (OpenAlex, retrieved 2026-09-28). Medium, used for all reported values: title or abstract contains (metasurface OR metasurfaces OR metamaterial OR metamaterials OR metalens) AND ("deep learning" OR "machine learning" OR "neural network" OR "neural networks"). Strict: metasurface terms and ("deep learning" OR "machine learning" OR "neural network" OR "inverse design") in the title. Broad: as medium, adding "meta-optics", "artificial intelligence" and "generative model". All three are restricted to the optical regime by AND (optical OR optics OR photonic OR photonics OR nanophotonic OR plasmonic OR light OR visible OR infrared OR laser OR "structural color" OR metalens OR metalenses OR nanostructure OR nanostructures) NOT (microwave OR terahertz OR THz OR GHz OR acoustic OR elastic OR mechanical OR antenna OR wireless OR "radio frequency"). Fig. 1a shading spans the three queries; 2026 is annualised as count × 365/271.

Counts. The medium query returns 1,430 works for 2019-2026; without the AI clause, 25,349. Against a random sample of 792 band-labelled works, the optical restriction had recall 0.90 and precision 0.96.

Experimental share. Of the 1,430 works, 191 have no abstract and 100 were labelled 'unclear', leaving 1,138. A language model (Claude) labelled each abstract 'yes' (fabricated, measured, or trained on own measured data) or 'no' (numerical only). The observed rate (34.2%) was corrected with the Rogan–Gladen estimator, $p = (p_{obs} + Sp - 1)/(Se + Sp - 1)$, with $Se = 0.911$ and $Sp = 0.979$ measured against the full-text labels of 92 full-text studies; 95% confidence intervals come from 5,000 bootstrap resamples. Corrected share: 36.0% (30.5–41.8%); 2019–2020, 19.2% (10–29%); 2021–2022, 30.3% (22–39%); 2023–2024, 41.6% (34–49%); 2025–2026, 38.2% (32–45%).

# Supplementary Table S1 | Full-text studies and their labels

One row per study (n = 100), ordered by year and OpenAlex citations. Terms are explained in Part 3.

### Part 1. Studies, data and models

| No. | First author | Year | Journal | Title | DOI | Citations | Band | Data | Model role | Physics in model | Model reused |
|---|---|---|---|---|---|---|---|---|---|---|---|
| 1 | An | 2019 | ACS Photonics | A Deep Learning Approach for Objective-Driven All-Dielectric Metasurface Design | 10.1021/acsphotonics.9b00966 | 421 | MIR (5–10 μm) | brute | surrogate, inverse (tandem) | yes | yes |
| 2 | Sakurai | 2019 | ACS Central Science | Ultranarrow-Band Wavelength-Selective Thermal Emission with Aperiodic Multilayered Metamaterials Designed by Bayesian Optimization | 10.1021/acscentsci.8b00802 | 180 | MIR (4–8 μm) | Bayesian | surrogate | no | no |
| 3 | Jiang | 2019 | Nanophotonics | Simulator-based training of generative neural networks for the inverse design of metasurfaces | 10.1515/nanoph-2019-0330 | 160 | NIR (800–1200 nm) | other | inverse (other) | yes | no |
| 4 | Liu | 2019 | Advanced Materials | Compounding Meta-Atoms into Metamolecules with Hybrid Artificial Intelligence Techniques | 10.1002/adma.201904790 | 146 | visible–NIR (500 nm–1.8 μm) | brute, curated | inverse (prior), surrogate | no | yes |
| 5 | Colburn | 2019 | Applied Optics | Optical frontend for a convolutional neural network | 10.1364/ao.58.003179 | 110 | visible (442/532/632 nm) | existing | co-design, classification | yes | no |
| 6 | Backer | 2019 | Optics Express | Computational inverse design for cascaded systems of metasurface optics | 10.1364/oe.27.030308 | 102 | visible (480–640 nm) | existing, other | classification | yes | no |
| 7 | Kiarashinejad | 2019 | Advanced Intelligent Systems | Knowledge Discovery in Nanophotonics Using Geometric Deep Learning | 10.1002/aisy.201900132 | 99 | visible (400–800 nm) | brute | inverse (other) | no | no |
| 8 | Lin | 2019 | Applied Optics | Achieving high numerical aperture near-infrared imaging based on an ultrathin cylinder dielectric metalens | 10.1364/ao.58.008914 | 14 | NIR (940 nm) | brute | surrogate | no | no |
| 9 | Harper | 2019 | 2019 IEEE Research and Applications of Photonics in Defense Conference (RAPID) | Machine Accelerated Nano-Targeted Inhomogeneous Structures | 10.1109/rapid.2019.8864295 | 5 | NOT STATED | brute | surrogate, inverse (direct) | no | no |
| 10 | Hemmatyar | 2019 | arXiv:1907.01595 (preprint) | Full color generation with Fano-type resonant HfO2 nanopillars designed by a deep-learning approach | 10.48550/arxiv.1907.01595 | 3 | visible (~400–700 nm) | brute | forward (other) | no | no |
| 11 | Chen | 2020 | Optics Express | Physics-informed neural networks for inverse problems in nano-optics and metamaterials | 10.1364/oe.384875 | 695 | NIR–MIR (2.1, 3 μm) | other | inverse (other) | yes | no |
| 12 | Kudyshev | 2020 | Applied Physics Reviews | Machine-learning-assisted metasurface design for high-efficiency thermal emitter optimization | 10.1063/1.5134792 | 232 | NIR (0.5–1.7 μm) | curated, augmented | inverse (prior), surrogate | no | no |
| 13 | Kiarashinejad | 2020 | npj Computational Materials | Deep learning approach based on dimensionality reduction for designing electromagnetic nanostructures | 10.1038/s41524-020-0276-y | 203 | NIR (1250–1850 nm) | brute, curated | surrogate, inverse (latent) | no | no |
| 14 | An | 2020 | Optics Express | Deep learning modeling approach for metasurfaces with high degrees of freedom | 10.1364/oe.401960 | 176 | MIR (5–10 μm) | brute | surrogate, inverse (GAN) | yes | no |
| 15 | Kudyshev | 2020 | Nanophotonics | Machine learning-assisted global optimization of photonic devices | 10.1515/nanoph-2020-0376 | 149 | visible–NIR (0.5–1.7 μm) | curated, augmented | inverse (prior), surrogate | no | no |
| 16 | Xu | 2020 | Advanced Photonics | Enhanced light-matter interactions in dielectric nanostructures via machine-learning approach | 10.1117/1.ap.2.2.026003 | 125 | NIR (1400–1600 nm) | brute | surrogate, inverse (tandem) | no | no |
| 17 | Rios | 2020 | Advanced Photonics Research | Multi-Level Electro-Thermal Switching of Optical Phase-Change Materials Using Graphene | 10.1002/adpr.202000034 | 125 | NIR–MIR (1550 nm, 2.21 μm) | other | inverse (other) | no (uncertain) | no |
| 18 | Wen | 2020 | ACS Photonics | Robust Freeform Metasurface Design Based on Progressively Growing Generative Networks | 10.1021/acsphotonics.0c00539 | 116 | visible–NIR (500–1300 nm) | active learning, augmented | inverse (GAN) | no | no |
| 19 | Lin | 2020 | Optics Letters | Inverse design of plasmonic metasurfaces by convolutional neural network | 10.1364/ol.387404 | 103 | visible–NIR (600–1600 nm) | brute, curated | inverse (direct) | no | no |

| No. | First author | Year | Journal | Title | DOI | Citations | Band | Data | Model role | Physics in model | Model reused |
|---|---|---|---|---|---|---|---|---|---|---|---|
| 20 | Moon | 2020 | Biosensors and Bioelectronics | Machine learning-based design of meta-plasmonic biosensors with negative index metamaterials | 10.1016/j.bios.2020.112335 | 103 | visible (632.8 nm) | brute | surrogate, inverse (other) | no | no |
| 21 | Yeung | 2020 | ACS Photonics | Elucidating the Behavior of Nanophotonic Structures through Explainable Machine Learning Algorithms | 10.1021/acsphotonics.0c01067 | 100 | MIR (4–12 μm) | brute | surrogate | yes | no |
| 22 | Ashalley | 2020 | Photonics Research | Multitask deep-learning-based design of chiral plasmonic metamaterials | 10.1364/prj.388253 | 99 | visible–NIR (400–1600 nm) | brute | surrogate, inverse (direct) | no | no |
| 23 | Pestourie | 2020 | npj Computational Materials | Active learning of deep surrogates for PDEs: application to metasurface design | 10.1038/s41524-020-00431-2 | 99 | optical (3 frequencies) | active learning | surrogate | no | yes |
| 24 | Yeung | 2020 | Nanophotonics | Multiplexed Supercell Metasurface Design and Optimization with Tandem Residual Networks | 10.1515/nanoph-2020-0549 | 98 | MIR–LWIR (4–12 μm) | brute, curated | inverse (tandem), surrogate | no | no |
| 25 | Han | 2020 | InfoMat | Inverse design of metasurface optical filters using deep neural network with high degrees of freedom | 10.1002/inf2.12116 | 89 | visible (400–680 nm) | brute, augmented | surrogate, inverse (tandem) | no | no |
| 26 | Wu | 2021 | Nature Communications | Programmable phase-change metasurfaces on waveguides for multimode photonic convolutional neural network | 10.1038/s41467-020-20365-z | 370 | NIR (1540–1580 nm) | existing | classification | no | no |
| 27 | Kulce | 2021 | Light: Science & Applications | All-optical information-processing capacity of diffractive surfaces | 10.1038/s41377-020-00439-9 | 157 | NOT STATED | existing, other | classification | yes | no |
| 28 | John-Herpin | 2021 | Advanced Materials | Infrared Metasurface Augmented by Deep Learning for Monitoring Dynamics between All Major Classes of Biomolecules | 10.1002/adma.202006054 | 124 | MIR (960–3060 cm-1) | measured | reconstruction | no | no |
| 29 | An | 2021 | Advanced Optical Materials | Deep Convolutional Neural Networks to Predict Mutual Coupling Effects in Metasurfaces | 10.1002/adom.202102113 | 123 | NIR (1.55 μm) | brute, augmented | surrogate | yes | no |
| 30 | Zhelyeznyakov | 2021 | ACS Photonics | Deep Learning to Accelerate Scatterer-to-Field Mapping for Inverse Design of Dielectric Metasurfaces | 10.1021/acsphotonics.0c01468 | 103 | visible (633, 400 nm) | curated | surrogate | yes | no |
| 31 | Deng | 2021 | Optics Express | Neural-adjoint method for the inverse design of all-dielectric metasurfaces | 10.1364/oe.419138 | 98 | IR (100–500 THz) | brute, active learning | surrogate | no | no |
| 32 | Xu | 2021 | Optical Materials Express | Efficient design of a dielectric metasurface with transfer learning and genetic algorithm | 10.1364/ome.427426 | 81 | NIR (1–3 μm) | brute, transfer | surrogate | no | yes |
| 33 | Yeung | 2021 | Advanced Optical Materials | Global Inverse Design across Multiple Photonic Structure Classes Using Generative Deep Learning | 10.1002/adom.202100548 | 78 | MIR (4–12 μm) | brute | inverse (GAN) | no | no |
| 34 | Lio | 2021 | Advanced Materials | Leveraging on ENZ Metamaterials to Achieve 2D and 3D Hyper-Resolution in Two-Photon Direct Laser Writing | 10.1002/adma.202008644 | 73 | visible–NIR (405–1000 nm) | other | inverse (other) | yes | no |
| 35 | Roberts | 2021 | Applied Physics Letters 119, 061101 | A deep learning approach to the forward prediction and inverse design of plasmonic metasurface structural color | 10.1063/5.0055733 | 71 | visible (400–750 nm) | brute, curated, augmented | surrogate, inverse (tandem) | yes | no |
| 36 | Lininger | 2021 | ACS Photonics | General Inverse Design of Layered Thin-Film Materials with Convolutional Neural Networks | 10.1021/acsphotonics.1c01498 | 67 | visible–NIR (450–950 nm) | brute, curated | inverse (direct), response translation | no | no |
| 37 | Wang | 2021 | Advanced Optical Materials | Visible Achromatic Metalens Design Based on Artificial Neural Network | 10.1002/adom.202101842 | 66 | visible (420–640 nm) | brute | surrogate | no | no |
| 38 | Jenkins | 2021 | Nanophotonics | Establishing exhaustive metasurface robustness against fabrication uncertainties through deep learning | 10.1515/nanoph-2021-0428 | 60 | NIR (1.55 μm) | brute | surrogate | yes | no |
| 39 | Seo | 2021 | ACS Photonics | Structural Optimization of a One-Dimensional Freeform Metagrating Deflector via Deep Reinforcement Learning | 10.1021/acsphotonics.1c00839 | 53 | NIR (900, 1000, 1100 nm) | active learning | inverse (RL) | yes | no |
| 40 | Garcia-Esteban | 2021 | Physical Review Applied | Deep learning for the modeling and inverse design of radiative heat transfer | 10.1103/physrevapplied.16.064006 | 49 | IR (thermal) | brute | surrogate | no | yes |
| 41 | Luo | 2022 | Light: Science & Applications | Metasurface-enabled on-chip multiplexed diffractive neural networks in the visible | 10.1038/s41377-022-00844-2 | 328 | visible (532 nm) | existing | classification | yes | yes |
| 42 | Ma | 2022 | Advanced Materials | Pushing the Limits of Functionality-Multiplexing Capability in Metasurface Design Based on Statistical Machine Learning | 10.1002/adma.202110022 | 241 | NIR (250–400 THz) | brute | surrogate, inverse (VAE) | no | yes |

| No. | First author | Year | Journal | Title | DOI | Citations | Band | Data | Model role | Physics in model | Model reused |
|---|---|---|---|---|---|---|---|---|---|---|---|
| 43 | Fan | 2022 | Nature Communications | Trilobite-inspired neural nanophotonic light-field camera with extreme depth-of-field | 10.1038/s41467-022-29568-y | 169 | visible (460–700 nm) | measured, augmented, physics-generated | reconstruction | no | no |
| 44 | Wang | 2022 | Nature Communications | Integrated photonic metasystem for image classifications at telecommunication wavelength | 10.1038/s41467-022-29856-7 | 124 | NIR (1490–1570 nm) | other | classification | yes | no |
| 45 | Zheng | 2022 | Science Advances | Meta-optic accelerators for object classifiers | 10.1126/sciadv.abo6410 | 82 | NIR (1.3 μm) | existing, measured | co-design, classification | yes | no |
| 46 | Chen | 2022 | APL Photonics | Physics-informed neural networks for imaging and parameter retrieval of photonic nanostructures from near-field data | 10.1063/5.0072969 | 76 | MIR (3 μm) | other | inverse (other) | yes | no |
| 47 | Patel | 2022 | Advanced Theory and Simulations 5, 2100604 | Ultra-Wideband, Polarization-Independent, Wide-Angle Multilayer Swastika-Shaped Metamaterial Solar Energy Absorber with Absorption Prediction using Machine Learning | 10.1002/adts.202100604 | 68 | UV–MIR (0.1–3 μm) | brute | surrogate | no | no |
| 48 | Parmar | 2022 | Scientific Reports | Graphene-based metasurface solar absorber design with absorption prediction using machine learning | 10.1038/s41598-022-06687-6 | 66 | UV–NIR (0.2–1.5 μm) | brute | surrogate | no | no |
| 49 | An | 2022 | Nanophotonics | Deep neural network enabled active metasurface embedded design | 10.1515/nanoph-2022-0152 | 63 | MIR (3–5 μm) | brute, model-generated | surrogate, inverse (direct) | no | no |
| 50 | Li | 2022 | Analytical Chemistry | Deep Learning-Based Multicapturer SERS Platform on Plasmonic Nanocube Metasurfaces for Multiplex Detection of Organophosphorus Pesticides in Environmental Water | 10.1021/acs.analchem.2c02973 | 63 | NIR (785 nm) | measured, augmented | classification | no | no |
| 51 | Wei | 2022 | Nanophotonics | End-to-end design of metasurface-based complex-amplitude holograms by physics-driven deep neural networks | 10.1515/nanoph-2022-0111 | 34 | NIR (1.5 μm) | existing, brute | inverse (tandem), surrogate, holography | yes | no |
| 52 | Zhang | 2022 | Optics Communications | End-to-end joint optimization of metasurface and image processing for compact snapshot hyperspectral imaging | 10.1016/j.optcom.2022.129154 | 34 | visible (460–700 nm) | existing, physics-generated | co-design, reconstruction | yes | no |
| 53 | Ma | 2022 | Advanced Optical Materials | Deep-Learning Enabled Multicolor Meta-Holography | 10.1002/adom.202102628 | 33 | visible (400–800 nm) | brute | surrogate | no | no |
| 54 | Yeung | 2022 | ACS Photonics | DeepAdjoint: An All-in-One Photonic Inverse Design Framework Integrating Data-Driven Machine Learning with Optimization Algorithms | 10.1021/acsphotonics.2c00968 | 27 | MIR (5–9 μm) | existing | inverse (GAN) | no | yes |
| 55 | Ma | 2022 | ACS Nano | Predicting Laser-Induced Colors of Random Plasmonic Metasurfaces and Optimizing Image Multiplexing Using Deep Learning | 10.1021/acsnano.2c02235 | 25 | visible | measured, brute | surrogate | no | no |
| 56 | He | 2023 | Opto-Electronic Advances | Pluggable multitask diffractive neural networks based on cascaded metasurfaces | 10.29026/oea.2024.230005 | 118 | NIR (800 nm) | existing, other | classification | yes | yes |
| 57 | Lin | 2023 | Nature Communications | Metasurface-empowered snapshot hyperspectral imaging with convex/deep (CODE) small-data learning theory | 10.1038/s41467-023-42381-5 | 115 | visible (500–650 nm) | measured, augmented | reconstruction | no | no |
| 58 | Chen | 2023 | Advanced Science 10, 2206718 | Broadband Solar Metamaterial Absorbers Empowered by Transformer-Based Deep Learning | 10.1002/advs.202206718 | 102 | solar (300–2500 nm) | brute | inverse (direct), surrogate | no | no |
| 59 | Kavungal | 2023 | Science Advances | Artificial intelligence-coupled plasmonic infrared sensor for detection of structural protein biomarkers in neurodegenerative diseases | 10.1126/sciadv.adg9644 | 90 | MIR (1500–1700 cm-1) | measured | reconstruction | no | no |
| 60 | Liu | 2023 | Advanced Functional Materials | Compatible Stealth Metasurface for Laser and Infrared with Radiative Thermal Engineering Enabled by Machine Learning | 10.1002/adfm.202212068 | 74 | NIR; MIR (3–15 μm) | brute, model-generated | surrogate, inverse (direct) | yes | no |
| 61 | Chen | 2023 | Light: Science & Applications | Neural network assisted high-spatial-resolution polarimetry with non-interleaved chiral metasurfaces | 10.1038/s41377-023-01337-6 | 70 | visible (470 nm) | measured, augmented | reconstruction | no | no |
| 62 | Chen | 2023 | Opto-Electronic Science | Chiral detection of biomolecules based on reinforcement learning | 10.29026/oes.2023.220019 | 59 | visible–NIR (~500–950 nm) | brute, active learning, Bayesian, augmented (uncertain) | surrogate | yes | no |

| No. | First author | Year | Journal | Title | DOI | Citations | Band | Data | Model role | Physics in model | Model reused |
|---|---|---|---|---|---|---|---|---|---|---|---|
| 63 | Kanmaz | 2023 | Optica | Deep-learning-enabled electromagnetic near-field prediction and inverse design of metasurfaces | 10.1364/optica.498211 | 58 | visible (400–700 nm) | brute | field predictor, inverse (direct) | no | yes |
| 64 | Wang | 2023 | Advanced Science | Customizing 2.5D Out-of-Plane Architectures for Robust Plasmonic Bound-States-in-the-Continuum Metasurfaces | 10.1002/advs.202206236 | 58 | NIR (1100–1400 nm) | brute | surrogate | no | no |
| 65 | Chen | 2023 | Advanced Optical Materials | All-Dielectric SERS Metasurface with Strong Coupling Quasi-BIC Energized by Transformer-Based Deep Learning | 10.1002/adom.202301697 | 54 | visible (532 nm) | other | surrogate, inverse (direct) | yes | no |
| 66 | Sarkar | 2023 | Advanced Photonics Research | Physics-Informed Machine Learning for Inverse Design of Optical Metamaterials | 10.1002/adpr.202300158 | 40 | visible (500–700 nm) | brute | inverse (tandem), surrogate | yes | no |
| 67 | Ueno | 2023 | Nanophotonics | Dual-band optical collimator based on deep-learning designed, fabrication-friendly metasurfaces | 10.1515/nanoph-2023-0329 | 28 | visible–NIR (650–780 nm) | brute | surrogate | no | no |
| 68 | Zhang | 2024 | Opto-Electronic Advances | Ultrahigh performance passive radiative cooling by hybrid polar dielectric metasurface thermal emitters | 10.29026/oea.2024.230194 | 131 | MIR (8–13 μm) | brute | surrogate | no | no |
| 69 | Zheng | 2024 | Nature Nanotechnology | Multichannel meta-imagers for accelerating machine vision | 10.1038/s41565-023-01557-2 | 114 | NIR (870 nm) | existing | classification | yes | no |
| 70 | Seo | 2024 | Advanced Photonics | Deep-learning-driven end-to-end metalens imaging | 10.1117/1.ap.6.6.066002 | 106 | visible (450/532/635 nm) | measured, existing, augmented | reconstruction | yes | no |
| 71 | Fan | 2024 | Nature Communications | Holographic multiplexing metasurface with twisted diffractive neural network | 10.1038/s41467-024-53749-6 | 104 | NOT STATED | physics-generated | holography | yes | no |
| 72 | Jiang | 2024 | Nature Communications 15, 8347 | Metasurface-enabled broadband multidimensional photodetectors | 10.1038/s41467-024-52632-8 | 84 | NIR–MIR (1–8 μm) | physics-generated | classification | no | no |
| 73 | Huang | 2024 | Nature Communications | Broadband thermal imaging using meta-optics | 10.1038/s41467-024-45904-w | 79 | LWIR (8–12 μm) | brute, curated | surrogate | no | no |
| 74 | Zhang | 2024 | Science Advances | Real-time machine learning-enhanced hyperspectro-polarimetric imaging via an encoding metasurface | 10.1126/sciadv.adp5192 | 71 | NIR (700–1150 nm) | measured, physics-generated | reconstruction | no | no |
| 75 | Meng | 2024 | Microsystems & Nanoengineering | Smart mid-infrared metasurface microspectrometer gas sensing system | 10.1038/s41378-024-00697-2 | 63 | MIR–LWIR (7–14 μm) | measured | classification | no | no |
| 76 | Dong | 2024 | ACS Photonics | Achromatic Single Metalens Imaging via Deep Neural Network | 10.1021/acsphotonics.3c01870 | 55 | visible (420–750 nm) | measured, existing, augmented | reconstruction | no | no |
| 77 | Ding | 2024 | ACS Materials Letters | Machine-Learning-Assisted Design of a Robust Biomimetic Radiative Cooling Metamaterial | 10.1021/acsmaterialslett.4c00337 | 55 | MIR (8–13 μm) | brute, augmented | surrogate, inverse (prior) | no | no |
| 78 | Gao | 2024 | Advanced Science | Meta-Attention Deep Learning for Smart Development of Metasurface Sensors | 10.1002/advs.202405750 | 49 | IR (~5–7 μm) | brute | surrogate, inverse (direct) | no | no |
| 79 | Wang | 2024 | Optical Materials | Inverse design for laser-compatible infrared camouflage metasurface enabled by physics-driven neural network and genetic algorithm | 10.1016/j.optmat.2024.115639 | 15 | NIR; IR (3–14 μm) | brute | surrogate | yes | no |
| 80 | Lee | 2024 | Advanced Optical Materials | Deep Neural Operator Enabled Concurrent Multitask Design for Multifunctional Metamaterials Under Heterogeneous Fields | 10.1002/adom.202303087 | 14 | optical (plasmonic) | curated | neural operator, field predictor | no | yes |
| 81 | Meng | 2025 | Science Advances | Ultranarrow-linewidth wavelength-vortex metasurface holography | 10.1126/sciadv.adt9159 | 86 | visible (413–700 nm) | other | holography | claimed only | no |
| 82 | Fu | 2025 | Optica | Miniaturized high-efficiency snapshot polarimetric stereoscopic imaging | 10.1364/optica.549864 | 42 | NIR (950–1350 nm) | physics-generated | reconstruction | yes | no |
| 83 | Elsayed | 2025 | Scientific Reports | Graphene metasurfaces biosensor for COVID-19 detection in the infra-red regime | 10.1038/s41598-025-92991-w | 39 | MIR (~4.2–6.0 μm) | brute | surrogate | no | no |
| 84 | Tang | 2025 | Nature Communications | Optical neural engine for solving scientific partial differential equations | 10.1038/s41467-025-59847-3 | 13 | visible (532 nm) | existing, brute | response translation, surrogate | yes | yes |
| 85 | Tao | 2025 | Optics Express | Deep learning-driven spectral exploration for metal-insulator-metal plasmonic metasurfaces | 10.1364/oe.565179 | 12 | visible–NIR (500–1000 nm) | measured, augmented | inverse (direct), surrogate, response translation | no | no |
| 86 | Zhang | 2025 | Nanophotonics | Chat to chip: large language model based design of arbitrarily shaped metasurfaces | 10.1515/nanoph-2025-0343 | 12 | NIR (1050–1600 nm) | brute | surrogate, inverse (other) | no | yes |

| No. | First author | Year | Journal | Title | DOI | Citations | Band | Data | Model role | Physics in model | Model reused |
|---|---|---|---|---|---|---|---|---|---|---|---|
| 87 | Riganti | 2025 | Advanced Optical Materials | Multiscale Physics-Informed Neural Networks for the Inverse Design of Hyperuniform Optical Materials | 10.1002/adom.202403304 | 11 | NIR–MIR (1.0–6.28 μm) | other | inverse (other) | yes | no |
| 88 | Lu | 2025 | ACS Photonics | An Agentic Framework for Autonomous Metamaterial Modeling and Inverse Design | 10.1021/acsphotonics.5c01514 | 11 | IR | existing, active learning | surrogate | no | yes |
| 89 | Liu | 2025 | Engineering Applications of Artificial Intelligence | A cascaded deep neural network for design and verification of surface lattice resonance metasurfaces biosensors | 10.1016/j.engappai.2025.110172 | 10 | visible (400–900 nm) | brute | surrogate, inverse (direct) | no | yes |
| 90 | Zhang | 2025 | ACS Photonics 12, 2070-2078 | High-Spatial and Spectral Resolution Snapshot Spectral Imaging Based on Nearfield Coupled Metasurfaces | 10.1021/acsphotonics.4c02504 | 10 | visible (400–700 nm) | existing, augmented | reconstruction | yes | no |
| 91 | Dong | 2025 | Nanoscale Horizons | An ultra-compact integrated phase shifter via electrically tunable meta-waveguides | 10.1039/d4nh00592a | 10 | NIR (1450–1650 nm) | other | classification | no | no |
| 92 | Lynch | 2025 | ACS Photonics | Physics-Guided Hierarchical Neural Networks for Maxwell's Equations in Plasmonic Metamaterials | 10.1021/acsphotonics.5c00552 | 9 | MIR (8–12 μm) | brute, physics-generated | surrogate | yes | yes |
| 93 | Shao | 2025 | Nanophotonics | Reliable, efficient, and scalable photonic inverse design empowered by physics-inspired deep learning | 10.1515/nanoph-2024-0504 | 9 | optical (on-chip photonics) | brute | surrogate | yes | yes |
| 94 | Ovcharenko | 2025 | Advanced Photonics Nexus | Forward and inverse design of single-layer metasurface-based broadband antireflective coating for silicon solar cells | 10.1117/1.apn.4.3.036009 | 9 | visible–NIR (500–1200 nm) | other | inverse (other) | yes | no |
| 95 | Zhang | 2025 | Laser & Photonics Reviews | Developing Neural Networks for Inverse Design of Chiral Metamaterials | 10.1002/lpor.202500051 | 8 | visible–NIR (400–1200 nm) | brute | surrogate, inverse (tandem) | no | no |
| 96 | Gladyshev | 2023 | Nanophotonics | Inverse design of all-dielectric metasurfaces with accidental bound states in the continuum | 10.1515/nanoph-2023-0373 | 21 | NIR (870 nm) | other | surrogate | yes | no |
| 97 | Liu | 2023 | Photonics Research | Self-design of arbitrary polarization-control waveplates via deep neural networks | 10.1364/prj.480845 | 19 | visible–NIR (650–950 nm) | brute | surrogate | no | yes |
| 98 | Lei | 2023 | Photonics Research | Dynamic multifunctional metasurfaces: an inverse design deep learning approach | 10.1364/prj.505991 | 18 | visible (520 nm) | brute | surrogate, inverse (VAE) | no | no |
| 99 | Zhang | 2024 | Nanophotonics | Snapshot computational spectroscopy enabled by deep learning | 10.1515/nanoph-2024-0328 | 13 | visible–NIR (400–900 nm) | brute, curated, measured, augmented | surrogate, classification | no | no |
| 100 | Behroozinia | 2024 | Nanophotonics | Leveraging multiplexed metasurfaces for multi-task learning with all-optical diffractive processors | 10.1515/nanoph-2024-0483 | 10 | visible (450/550/650 nm) | existing, brute | classification, surrogate | yes | yes |

**Part 2. Measurement, design and cost**

| No. | First author | Year | Measured | Fabrication-aware | Measurement fed back | Coupling | Task objective | Non-AI comparison | Data-generation time | Hardware | Training time |
|---|---|---|---|---|---|---|---|---|---|---|---|
| 1 | An | 2019 | no | no | no | no | no | cited only | NOT STATED | NOT STATED | NOT STATED |
| 2 | Sakurai | 2019 | yes | no | no | no | no | quantified | ~24 d (incl. optimization) | 24-core cluster, 128 GB | NOT STATED |
| 3 | Jiang | 2019 | no | no | no | no | no | quantified | N/A (in-loop simulation) | Titan V GPU, 4 CPUs | 10 min per device |
| 4 | Liu | 2019 | yes | no | no | no | no | cited only | NOT STATED | single Quadro P5000 GPU | NOT STATED |
| 5 | Colburn | 2019 | no | no | no | no | yes | none | N/A (public dataset) | i7 8-core CPU; TitanX GPU | NOT STATED |
| 6 | Backer | 2019 | no | yes | no | no | yes | none | ~minutes per meta-atom (FDTD) | NOT STATED | NOT STATED |
| 7 | Kiarashinejad | 2019 | yes | no | no | no | no | cited only | NOT STATED | PC, Core i7-6700, 16 GB RAM | NOT STATED |
| 8 | Lin | 2019 | yes | no | no | reduced only | no | cited only | NOT STATED | NOT STATED | NOT STATED |
| 9 | Harper | 2019 | no | no | no | no | no | per-inference only | NOT STATED | GTX 1080 Ti GPUs | NOT STATED |
| 10 | Hemmatyar | 2019 | yes | no | no | no | no | cited only | NOT STATED | NOT STATED | NOT STATED |
| 11 | Chen | 2020 | no | no | no | no | no | run, not quantified | NOT STATED | NOT STATED | NOT STATED |
| 12 | Kudyshev | 2020 | no | yes | no | no | no | quantified | 1.64 h per TO design | 2× 12-core Xeon Gold, 96 GB | NOT STATED |
| 13 | Kiarashinejad | 2020 | yes | no | no | no | no | cited only | NOT STATED | PC, Core i7-6700, 8 GB RAM | <3 h (forward + inverse) |
| 14 | An | 2020 | no | no | no | no | no | cited only | NOT STATED | NOT STATED | NOT STATED |
| 15 | Kudyshev | 2020 | no | no | no | no | no | cited only | 328 h (600 TO designs) | 2× 12-core Xeon Gold, 96 GB RAM | 35 min; 48 min |
| 16 | Xu | 2020 | yes | no | no | no | no | cited only | NOT STATED | Core i7-4770, 16 GB (inference) | NOT STATED |
| 17 | Rios | 2020 | no | no | no | no | no | none | NOT STATED | NOT STATED | NOT STATED |
| 18 | Wen | 2020 | no | yes | no | no | no | quantified | NOT STATED | Sherlock cluster (simulations) | NOT STATED |
| 19 | Lin | 2020 | yes | no | no | no | no | cited only | NOT STATED | NOT STATED | NOT STATED |
| 20 | Moon | 2020 | yes | no | no | no | no | quantified | NOT STATED | NOT STATED | NOT STATED |
| 21 | Yeung | 2020 | no | no | no | no | no | per-inference only | ~30 min per FDTD simulation | Core i5-8600T CPU | 340 min (selected CNN) |
| 22 | Ashalley | 2020 | no | no | no | no | no | cited only | >3 h per spectrum; 80 d (incl. training) | i7-CPU PC (330 W) | NOT STATED |
| 23 | Pestourie | 2020 | no | no | no | no | no | per-inference only | 27.8k s (128k points) | 3.5 GHz 6-core Xeon E5 | 5.4k s (incl. evaluation) |
| 24 | Yeung | 2020 | no | no | no | learned | no | cited only | ~75 d total; ~30 min per simulation | HPC cluster, 4 nodes per simulation | NOT STATED |
| 25 | Han | 2020 | no | no | no | no | no | per-inference only | 78,992 ms per structure (RCWA) | i5-6300HQ; GTX 960M (timing only) | NOT STATED |
| 26 | Wu | 2021 | yes | no | no | no | yes | none | N/A (public dataset) | NOT STATED | NOT STATED |
| 27 | Kulce | 2021 | no | no | no | no | yes | none | N/A (physics in loop) | NOT STATED | NOT STATED |
| 28 | John-Herpin | 2021 | yes | no | yes | no | no | run, not quantified | NOT STATED | NOT STATED | NOT STATED |
| 29 | An | 2021 | no | no | no | learned | no | quantified | NOT STATED | NOT STATED | NOT STATED |
| 30 | Zhelyeznyakov | 2021 | no | no | no | learned | no | cited only | ~16 h (10 FDTD simulations) | NOT STATED | NOT STATED |
| 31 | Deng | 2021 | no | no | no | learned | no | cited only | NOT STATED | NOT STATED | NOT STATED |
| 32 | Xu | 2021 | no | no | no | no | no | cited only | NOT STATED | PC, RTX 2060 GPU | ~150 min per 10,000 epochs |
| 33 | Yeung | 2021 | no | no | no | no | no | cited only | ~65 d total; ~5 min per simulation | HPC cluster; Titan RTX GPU | ~30 min (final model) |
| 34 | Lio | 2021 | yes | yes | no | no | no | none | N/A | NOT STATED | NOT STATED |
| 35 | Roberts | 2021 | no | no | no | no | no | per-inference only | NOT STATED | NOT STATED | NOT STATED |
| 36 | Lininger | 2021 | no | no | no | no | no | quantified | NOT STATED | HPC node, 10 cores, 64 GB, 2080 GPU | ~3 h (300 epochs) |
| 37 | Wang | 2021 | yes | no | no | no | no | cited only | ~116 h (695 CST simulations) | Core i7-6560U laptop; Xeon E5-2667 v4 | <10 s per shape (30 s total) |
| 38 | Jenkins | 2021 | no | yes | no | learned | no | run, not quantified | 3.16 d (81,408 samples) | 4× RTX 2080 Ti (44 GB) | 3.15 d |
| 39 | Seo | 2021 | no | no | no | no | no | quantified | NOT STATED | NOT STATED | NOT STATED |
| 40 | Garcia-Esteban | 2021 | no | no | no | no | no | cited only | up to ~24 h per spectrum | 2.3-GHz Xeon desktop, 18 CPUs | NOT STATED |
| 41 | Luo | 2022 | yes | no | no | no | yes | none | N/A (public datasets) | RTX 2080 Ti, i9-10980XE, 128 GB | few min to tens of min per network |

| No. | First author | Year | Measured | Fabrication-aware | Measurement fed back | Coupling | Task objective | Non-AI comparison | Data-generation time | Hardware | Training time |
|---|---|---|---|---|---|---|---|---|---|---|---|
| 42 | Ma | 2022 | yes | no | no | no | no | cited only | NOT STATED | NOT STATED | NOT STATED |
| 43 | Fan | 2022 | yes | no | yes | no | no | cited only | NOT STATED | NOT STATED | NOT STATED |
| 44 | Wang | 2022 | yes | yes | no | reduced only | yes | none | N/A (physics in loop) | NOT STATED | NOT STATED |
| 45 | Zheng | 2022 | yes | yes | yes | no | yes | none | NOT STATED | Quadro RTX 5000 | NOT STATED |
| 46 | Chen | 2022 | no | no | no | no | no | cited only | NOT STATED | i7-8700K, 32 GB, GTX 1080Ti | ~10 h per PINN training |
| 47 | Patel | 2022 | no | no | no | no | no | quantified | NOT STATED | NOT STATED | NOT STATED |
| 48 | Parmar | 2022 | no | no | no | no | no | cited only | NOT STATED | NOT STATED | NOT STATED |
| 49 | An | 2022 | no | no | no | no | no | per-inference only | minutes per meta-atom (single CPU) | 16-core CPU; 1080Ti GPU | 72 h (forward DNNs) |
| 50 | Li | 2022 | yes | no | yes | no | no | cited only | 5 s per SERS spectrum | NOT STATED | NOT STATED |
| 51 | Wei | 2022 | no | no | no | no | no | quantified | NOT STATED | Xeon Silver 4210; RTX 2080 Ti; 30 GB | ~2 h (Y-Net) |
| 52 | Zhang | 2022 | no | no | no | no | yes | quantified | N/A (public ICVL dataset) | i9-10900K; RTX 3090 (24 GB) | NOT STATED |
| 53 | Ma | 2022 | yes | yes | no | no | no | cited only | NOT STATED | GTX 1660 Ti (6 GB) | ~30 min per network |
| 54 | Yeung | 2022 | no | yes | no | simulated | no | quantified | NOT STATED | HPC cluster, 4 nodes (Hoffman2) | NOT STATED |
| 55 | Ma | 2022 | yes | yes | yes | no | no | run, not quantified | NOT STATED | GTX 1080Ti cluster GPUs | 2 weeks per mode (six modes) |
| 56 | He | 2023 | yes | no | no | no | yes | none | N/A (public datasets) | Core i5-10500, 32 GB RAM | NOT STATED |
| 57 | Lin | 2023 | yes | no | yes | no | no | run, not quantified | NOT STATED | NOT STATED | NOT STATED |
| 58 | Chen | 2023 | yes | no | no | no | no | cited only | NOT STATED | GTX 3080Ti, i7-10700K, 16 GB | NOT STATED |
| 59 | Kavungal | 2023 | yes | no | yes | no | no | cited only | NOT STATED | NOT STATED | NOT STATED |
| 60 | Liu | 2023 | yes | no | no | no | no | cited only | ~40 d (8000 FDTD simulations) | NOT STATED | NOT STATED |
| 61 | Chen | 2023 | yes | no | yes | no | no | run, not quantified | NOT STATED | NOT STATED | NOT STATED |
| 62 | Chen | 2023 | yes | no | no | no | no | cited only | NOT STATED | NOT STATED | NOT STATED |
| 63 | Kanmaz | 2023 | no | no | no | learned | no | cited only | NOT STATED | NOT STATED | NOT STATED |
| 64 | Wang | 2023 | yes | no | no | no | no | per-inference only | NOT STATED | i9-12900K, 64 GB (FDTD) | NOT STATED |
| 65 | Chen | 2023 | no | no | no | no | no | cited only | NOT STATED | NOT STATED | NOT STATED |
| 66 | Sarkar | 2023 | no | no | no | no | no | none | NOT STATED | Pittsburgh Supercomputing Center (ACCESS/XSEDE) | NOT STATED |
| 67 | Ueno | 2023 | yes | yes | no | no | no | run, not quantified | NOT STATED | NOT STATED | NOT STATED |
| 68 | Zhang | 2024 | yes | no | no | no | no | cited only | NOT STATED | NOT STATED | NOT STATED |
| 69 | Zheng | 2024 | yes | no | no | no | yes | none | N/A (public datasets) | Quadro RTX 5000 GPU | NOT STATED |
| 70 | Seo | 2024 | yes | no | yes | no | no | cited only | NOT STATED | RTX 4090 24 GB | NOT STATED |
| 71 | Fan | 2024 | yes | yes | no | no | no | none | N/A | NOT STATED | NOT STATED |
| 72 | Jiang | 2024 | yes | no | no | no | no | none | NOT STATED | NOT STATED | NOT STATED |
| 73 | Huang | 2024 | yes | no | no | no | no | quantified | NOT STATED | NOT STATED | NOT STATED |
| 74 | Zhang | 2024 | yes | no | yes | reduced only | no | quantified | NOT STATED | i9-9900K, RTX 2080 Ti, 32 GB | ~12 h |
| 75 | Meng | 2024 | yes | no | yes | no | no | cited only | 5 s readout interval per sample | NOT STATED | NOT STATED |
| 76 | Dong | 2024 | yes | no | yes | no | no | cited only | NOT STATED | 2× RTX 4090 | NOT STATED |
| 77 | Ding | 2024 | yes | no | no | no | no | quantified | NOT STATED | NOT STATED | NOT STATED |
| 78 | Gao | 2024 | yes | no | no | no | no | cited only | NOT STATED | GTX 3080Ti, i7-10700K, 16 GB | ~5 h per network |
| 79 | Wang | 2024 | no | no | no | no | no | per-inference only | NOT STATED | NOT STATED | NOT STATED |
| 80 | Lee | 2024 | no | no | no | learned | no | cited only | NOT STATED | Tesla T4 GPU, 16 GB | NOT STATED |
| 81 | Meng | 2025 | yes | no | no | no | no | run, not quantified | NOT STATED | Titan RTX 24GB GPU | NOT STATED |
| 82 | Fu | 2025 | yes | no | no | no | no | quantified | NOT STATED | NOT STATED | NOT STATED |
| 83 | Elsayed | 2025 | no | no | no | no | no | none | NOT STATED | NOT STATED | NOT STATED |
| 84 | Tang | 2025 | yes | yes | yes | no | yes | cited only | NOT STATED | RTX 6000 GPU | NOT STATED |

| No. | First author | Year | Measured | Fabrication-aware | Measurement fed back | Coupling | Task objective | Non-AI comparison | Data-generation time | Hardware | Training time |
|---|---|---|---|---|---|---|---|---|---|---|---|
| 85 | Tao | 2025 | yes | no | yes | no | no | run, not quantified | NOT STATED | RTX 4060Ti 16 GB | LSTM 4.43–5.28 h; CNN 2.74–3.11 min |
| 86 | Zhang | 2025 | no | no | no | no | no | per-inference only | NOT STATED | 2× Xeon Gold 6258R, 1.5 TB; RTX 2080 Ti | NOT STATED |
| 87 | Riganti | 2025 | no | yes | no | learned | no | cited only | NOT STATED | P100, V100 or A40 GPUs | NOT STATED |
| 88 | Lu | 2025 | no | no | no | no | no | none | avoided (existing dataset subsampled) | GTX 3090 GPU | 564 s; 2073 s |
| 89 | Liu | 2025 | no | no | no | no | no | per-inference only | 14 d (16,236 samples, ~1 min each) | i7-12700H laptop, RTX 3070 Ti, 16 GB | 163 s (SPMLP forward model) |
| 90 | Zhang | 2025 | yes | yes | no | no | no | cited only | NOT STATED | 1× RTX 4090 GPU | NOT STATED |
| 91 | Dong | 2025 | yes | no | no | no | no | none | N/A | NOT STATED | NOT STATED |
| 92 | Lynch | 2025 | no | no | no | no | no | cited only | NOT STATED | NOT STATED | NOT STATED |
| 93 | Shao | 2025 | no | no | no | learned | yes | quantified | NOT STATED | NOT STATED | NOT STATED |
| 94 | Ovcharenko | 2025 | no | uncertain | no | no | no | quantified | 6–7 min (parametric sweep) | 16-core CPU, no GPU | <1 h to several h |
| 95 | Zhang | 2025 | yes | no | no | no | no | per-inference only | 9 min per structure (FDTD) | NOT STATED | NOT STATED |
| 96 | Gladyshev | 2023 | no | uncertain | no | no | no | cited only | NOT STATED | NOT STATED | NOT STATED |
| 97 | Liu | 2023 | yes | yes | no | no | no | per-inference only | NOT STATED | NOT STATED | NOT STATED |
| 98 | Lei | 2023 | no | no | no | no | no | none | NOT STATED | NOT STATED | NOT STATED |
| 99 | Zhang | 2024 | yes | yes | yes | no | no | cited only | NOT STATED | NOT STATED | NOT STATED |
| 100 | Behroozinia | 2024 | no | no | no | no | yes | none | NOT STATED | V100 32 GB (Colab Pro) | NOT STATED |

**Part 3. Terms**

| Column | Entry | Meaning |
|---|---|---|
| General | NOT STATED | Not reported |
| General | N/A | No such stage |
| General | uncertain | Ambiguous in the text; excluded where Table S2 says so |
| General | other | Fits none of the listed categories |
| Data | brute | Predetermined sweep or random sampling |
| Data | augmented | Generative or geometric augmentation |
| Data | physics-generated | Data from a physical or analytic model |
| Data | model-generated | Data from a learned model |
| Data | transfer | Pretrained model reduces the data needed |
| Model role | surrogate | Predicts the response from the structure |
| Model role | response translation | Maps one response to another |
| Model role | inverse (...) | Inverse design; network type in brackets |
| Model role | reconstruction / classification / holography | Computational back end |
| Model role | co-design | Optics and back end optimized jointly |
| Physics in model | yes | Physics in the architecture, the loss, or a differentiable solver |
| Physics in model | claimed only | Physics claimed but not shown |
| Model reused | yes | A trained model reused for another task or serving several tasks |
| Coupling | simulated / learned / reduced only | Suppl. Methods 4 |
| Task objective | yes | Suppl. Methods 4 |
| Non-AI comparison | quantified / run, not quantified / per-inference only / cited only / none | Suppl. Methods 3 |

## Supplementary Table S2 | Number ledger

Every number in the main text, Box 1 and the figures. Full-text values are counts of the 100 studies (equal to percentages). 'Rule' gives the Table S1 column and entry counted, or the method; for values taken from a cited study, 'Source' gives a verbatim excerpt (… marks an omission).

| Location | Statement | Value | Rule | Source |
|---|---|---|---|---|
| Summary; Intro; Box 1 | Full-text studies | 100 (15/yr; 2019: 10) | Suppl. Methods 2 | Table S1 |
| Box 1 | Optical metasurface works, 2019–2026 | 25,349 | Medium query without AI clause | Suppl. Methods 5 |
| Intro; Box 1 | AI-related optical works | 1,430 (~1,400) | Medium query | Suppl. Methods 5 |
| Box 1; Fig. 1 | Abstract-labelled works | 1,138 | 1,430 − 191 no abstract − 100 unclear | Suppl. Methods 5 |
| Intro; Box 1; Fig. 1a | Experimental share, field-wide | 36.0% (~36%) | Rogan–Gladen corrected | Suppl. Methods 5 |
| Section 2; Fig. 1a | Experimental share by period | 19.2%; 30.3%; 41.6%; 38.2% | Two-year bins, 2019–2026 | Suppl. Methods 5 |
| Section 2 | AI share of optical literature | 5.6%; 8.5% (2025–2026) | 1,430/25,349; 691/8,177 | Suppl. Methods 5 |
| Fig. 1a | 2026, annualised | 469 (~470) | 348 × 365/271 | Suppl. Methods 5 |
| Intro; Fig. 1b | Compare with a non-AI workflow | 27 (not: 73) | Non-AI comparison: quantified or run, not quantified | Table S1 |
| Intro | Quantify both workflows | 18 | Non-AI comparison: quantified | Table S1 |
| Intro; Fig. 1b | Data-generation time recorded | 34 (not: 66) | Data-generation time stated | Table S1 |
| Fig. 1b | Complete cost record | 12 (not: 88) | Hardware, data-generation and training time all stated | Table S1 |
| Intro; Shift 2; Fig. 2 | Reuse existing data | 17 | Data: existing | Table S1 |
| Intro; Fig. 1b, 1c | Include measurements | 50 (simulation only: 50) | Measured: yes | Table S1 |
| Intro; Shift 3; Fig. 1c, 2 | Measurements fed back into design | 15 | Measurement fed back: yes | Table S1 |
| Shift 3; Fig. 1c, 2 | Fabrication-aware design | 17 | Fabrication-aware: yes | Table S1 |
| Section 2 | Sweep or random sampling | 54 | Data: brute | Table S1 |
| Section 2; Fig. 1c | Surrogate model | 57 | Model role: surrogate | Table S1 |
| Fig. 1c | Inverse design | 42 | Model role: any inverse | Table S1 |
| Shift 1 | No physics in the model | 60 | Physics in model: no | Table S1 |
| Fig. 2 | Embed physics | 38 | Physics in model: yes | Table S1 |
| Fig. 2 | Reuse a model or data; share a model; cross band | 30; 19; 0 | Data: existing, or Model reused: yes; Model reused: yes; no reuse across bands | Table S1 |
| Fig. 2 | Programmable hardware | 1 | Title keyword | Table S1 |
| Shift 4; Fig. 2 | Coupling computed or learned | 10 (1 + 9); reduced only: 3 | Coupling: simulated or learned; reduced only | Table S1 |
| Shift 5; Fig. 2 | Task-level design objective | 13 | Task objective: yes | Table S1 |
| Section 2 | MetasurfaceViT dataset | > 60 million | As reported | Yan 2026 (ref. 41): "expand the Jones matrices training dataset to 60 million samples" [p. 2] |
| Section 3 (T1) | Full-wave simulation limit | 12 × 12 | As reported | Rozman 2024 (ref. 44): "arrays up to as large as 12 × 12 elements … within a practical time frame" [p. 3] |

| Location | Statement | Value | Rule | Source |
|---|---|---|---|---|
| Section 3 (T1) | Coupling-aware design scale | cm; 24 × 24; 224 μm | As reported | Xu 2026 (ref. 20): "micrometer to centimeter scales" [Abstract]; "24 × 24 nanofin configuration … FDTD-computed results" [Fig. 5]; "224 × 224 μm" [p. 7] |
| Section 3 (T3) | Training vs parameter sweep | a few hours; 6–7 min | As reported | Ovcharenko 2025 (ref. 48): "a few hours of training time, which is several times longer than a more precise parametric sweep" [p. 7]; "~6 to 7 min … on a 16-core processor" [p. 7] |
| Shift 2; Fig. 1c | Simulation share of time | 85%; 99.8% | As reported | Pestourie 2020 (ref. 9): "85% of the total computation time" [p. 6] \| Wei 2026 (ref. 45): "18.80 h (dataset generation: 18.79 h" [p. 8] |
| Shift 2 | Data campaigns | 14 days; 5 months | As reported | Liu 2025 (ref. 55): "over a period of 14 days" [p. 6] \| Lin 2025 (ref. 54): "over a period of 5 months" [p. 11] |
| Shift 2 | Target samples after pretraining | 1,800 | As reported | Zhu 2025 (ref. 57): "public data set" [p. 2]; "merely 1800 samples" [Abstract] |
| Shift 3 | Metalens efficiency | 64%; 69% | As reported | Liao 2022 (ref. 71): "measured focusing efficiency of 64% … calculated value of 69%" [Abstract] |
| Shift 3 | Fabrication rounds; insertion loss | 6; 3.24 vs 8.8 dB | As reported | Witt 2023 (ref. 72): "Six rounds" [Sec. V]; "3.24 dB … 8.8 dB" [Sec. VII] |